\documentstyle[11pt,epsfig]{article}
\begin{document}
\title{Projectile fragment emission in fragmentation of $^{56}$Fe on C, Al, and CH$_{2}$ targets at 471 A MeV
\thanks{Submitted to Chin. Phys. C}}

\author{Yan-Jing Li$^{a}$, Dong-Hai Zhang$^{b}$\thanks{Corresponding author. Tel: +863572051347; fax: +863572051347. E-mail address:zhangdh@dns.sxnu.edu.cn}, Shiwei Yan$^{a,c}$
, Li-Chun Wang$^{a}$,\\ Jin-Xia Cheng$^{a}$, Jun-Sheng Li$^{b}$, S. Kodaira$^{d}$, and N. Yasuda$^{d}$ \\
$^{a}$College of Nuclear Science and Technology, Beijing Normal University, Beijing 100875, China\\
$^{b}$Institute of Modern Physics, Shanxi Normal University, Linfen 041004, China\\
$^{c}$Beijing Radiation Center, Beijing 100875, China\\
$^{d}$Fundamental Technology Center, National Institute of Radiological Sciences\\
4-9-1 Anagawa, Inage-ku, Chiba 263-8555, Japan}

\date{}
\maketitle

\begin{center}
\begin{minipage}{140mm}
\vskip 0.4in
\begin{center}{\bf Abstract}\end{center}
{The emission angle and the transverse momentum distributions of projectile fragments produced in fragmentation of $^{56}$Fe on CH$_{2}$, C, and Al targets at 471 A MeV are measured. It is found that for the same target the average value and width of angular distribution decrease with increase of the projectile fragment charge, and for the same projectile fragment the average value of the distribution increases and the width of the distribution decreases with increasing the target charge number. The transverse momentum distribution of projectile fragment can be explained by a single Gaussian distribution and the averaged transverse momentum per nucleon decreases with the increase of the charge of projectile fragment. The cumulated squared transverse momentum distribution of projectile fragment can be well explained by a single Rayleigh distribution. The temperature parameter of emission source of projectile fragment, calculated from the cumulated squared transverse momentum distribution, decreases with the increase of the size of projectile fragment.}\\

\end{minipage}
\end{center}

\vskip 0.4in
\baselineskip 0.2in
\section{Introduction}

The knowledge of heavy ion fragmentation at intermediate and high energy is very important in nuclear physics, astrophysics, and medical physics. Considering the biological effects of space radiation, when astronaut have their mission outside the earth magnetic field, they are suffered from Galactic Cosmic Radiation(GCR) and Solar particle events, {\em e.g.}, showers of energetic charged particles from the surface of the Sun. These energetic charged particles will be the dominant sources of the radiation dose and affect the health of humans on long-duration spaceflight both inside and outside the station. According to the GCR model developed by Badhwar and O'Neill\cite{ref1}, in unshielded free space in the inner heliosphere, iron ions deliver about $8\%$ of the total dose from the GCR and $27\%$ of the dose equivalent at times near solar maximum, even though they contribute less than $1\%$ of the total GCR flux. Because iron ions are the most densely ionizing particles which are presented in significant numbers in the GCR, there has been considerable interest in understanding their transport through matter and their biological effects.

The understanding of the radiobiology of heavy charged particles (HZE) is a subject of great interest due to the complicated dependence of their relative biological effectiveness on the type of ion and its energy, and its interaction with various targets. It has become clear that heavy ions have the largest radiological effects. These effects also appear in regions close to the beam entrance, {\it i.e.}, in the depth-dose plateau region, where normal tissue is usually situated. In addition, due to the longer ranges of the fragments produced by the fragmentation of the incident ions, the tails of the dose distribution beyond the Bragg peak may be too high for minimizing doses to normal tissue beyond the primary ion range. Finally, recent experimental results\cite{ref2} have revealed that the fragments are emitted at larger angles than the scattering angles of the beam, which further increases the spread of the beam. Exact information about the fragment emission angular distributions will be especially important in radiotherapy. So far only a few measurements have been performed to analyze fragment emission angles from HZE reactions below 500 A MeV[2-4].

Fragmentation is a term commonly used to specify a nuclear disassembly by force. Hot fragmentation is meant to indicate the most violent of these process, following excitation beyond the limits of nuclear binding, but still ending with bound nuclear fragments of different sizes in the final channels\cite{ref5}. The formation mechanism of these fragments, whether they the remnants of an incomplete destruction or the products of a condensation out of the disordered matter, has continued to be the topic of very active research. In order to describe the physical process of heavy ion transport, several one-dimensional Monte Carlo codes, such as HZETRN\cite{ref6}, HIBRAC\cite{ref7}, FLUKA\cite{ref8}, NUCFRAG2\cite{ref9}, and three dimensional model\cite{ref10} are appeared. The improved quantum molecular dynamics model (ImQMD) is a dynamical model which is developed to follow the reaction process on a microscopic level\cite{ref11,ref12}.

The properties of $^{56}$Fe on various targets at various energies have been studied by many groups[13-25], most of the studies are devoted to the total charge-changing cross sections and the partial cross sections of fragment productions, a little attention is paid to the fragment emission angular distribution study.

In this paper, we present the results of the emission angular distribution, transverse momentum distribution and the temperature of emission source of fragment produced in fragmentation of 471 A MeV $^{56}$Fe on C, Al, and CH$_{2}$ targets. CH$_{2}$ target is used to obtain the cross section on a hydrogen target. The fragmentation cross sections is published in our previous paper\cite{ref26}. The arrangements of this paper are as follows: we introduce our experimental detail in sec. II. In sec. III, experiment result and discussion are given. At last, we give the conclusion in sec. IV.

\section{Experimental details}

\subsection{Experimental setup}
Stacks of C, Al and CH$_{2}$ targets sandwiched with CR-39 detectors were exposed normally to 471 A MeV $^{56}$Fe beams at the Heavy Ion Medical Accelerator in Chiba (HIMAC) at the Japanese National Institute of Radiological Sciences (NIRS).  Figure 1 shows the configuration of sandwiched target. A CR-39 sheet, about 0.77 mm in thickness, is placed before and after the targets. The thickness of carbon, aluminum and polyethylene targets is  5, 3, and 10 mm, respectively.
\begin{figure}[htbp]
\begin{center}
\includegraphics[width=0.70\linewidth]{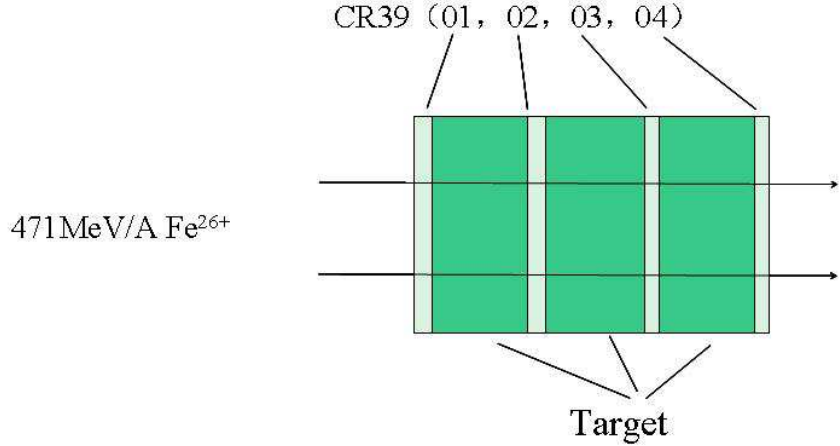}
\caption{Sketch of the target-detector configuration.}
\end{center}
\end{figure}

After exposure, the CR-39 detectors are etched in 7N NaOH aqueous solution at temperature of 70$^{\circ}$C for 15 hours. Then, the beam ions and their fragments manifest in the CR-39 as etch-pit cones on both sides of CR-39 sheets. The images of ion tracks are scanned and analyzed automatically by HSP-1000 microscope system and the PitFit track measurement software, then checked manually. The PitFit software allows us to extract some geometric information, such as the position coordinates, major and minor axes and area of etched track spot on CR-39 surfaces. Image data ($45\times45mm^{2}$) are acquired for both front and back surfaces of each CR-39 detector. About 2$\times$10$^{4}$ Fe ions are traced from the first CR-39 detector surface in the stack. $^{56}$Fe trajectories and the ones of secondary fragments are reconstructed in the whole stack.

\subsection{Experimental method}
The spots on the front surface (with respect to the beam direction) are directly scanned firstly, then the CR-39 sheet is turned under the middle line of the sheet and the spots on the back surface are scanned. The trajectories of ion tracks through CR-39 sheets are reconstructed in two steps using the track tracing method\cite{ref27}: (1) the track position in CR-39 surfaces is corrected by parallel and rotational coordinate translation (except for the track position on upper surface of the first CR-39 sheet), and (2) the difference between the track position of corresponding tracks on both side of the CR-39 sheets and on the surfaces neighboring targets is minimized by a track matching routine. The coordinate of track before the target (or front surface of CR-39 sheet) is $(x,y)$ and of matching track after the target (or back surface of CR-39 sheet) is $(x',y')$. Following the translation relation, the coordinate of matching track can be calculated as:
\begin{eqnarray}
 x'_{th}= & ax + by+c,
 \end{eqnarray}
\begin{eqnarray}
 y'_{th}= & a'x+b'y+c',
\end{eqnarray}
parameters $a, b, c, a', b'$, and $c'$ are determined using the least square method. Then, the coordinate $x'_{th}$, $y'_{th}$ of matching track is calculated. However, because of the Coulomb scatter etc. contributions, $x'_{th}$, $y'_{th}$ are certainly different from $x', y'$, the difference $dx=x'_{th}-x'$, $dy=y'_{th}-y'$ is calculated which can help us to determine the matching track.

\begin{figure}[htbp]
\begin{center}
\includegraphics[width=0.70\linewidth]{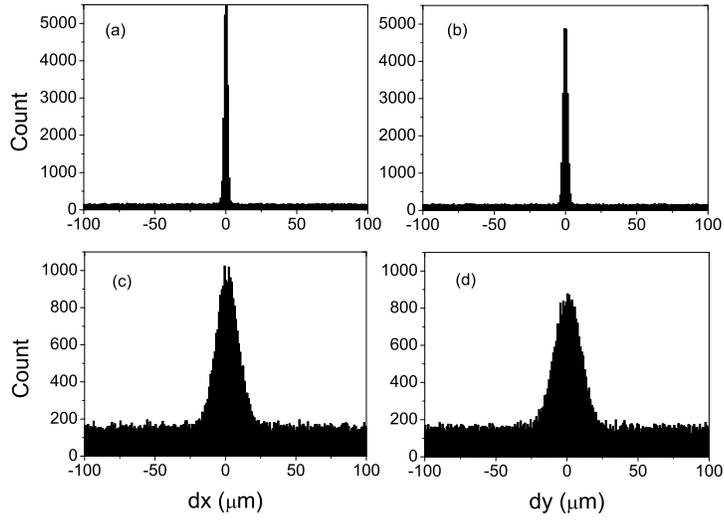}
\caption{The distribution of dx and dy. (a) and (b) the difference between the front and back surface on a CR-39 sheet, (c) and (d) the difference before and after carbon target.}
\end{center}
\end{figure}

\begin{figure}[htbp]
\begin{center}
\includegraphics[width=0.70\linewidth]{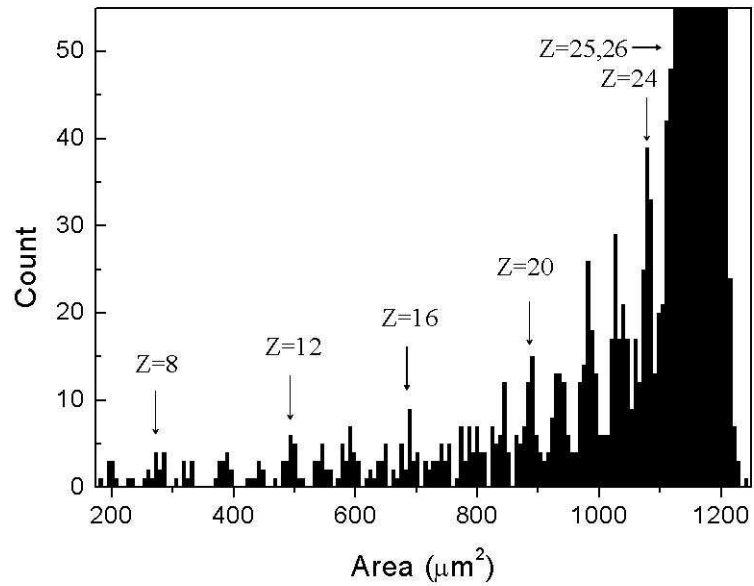}
\caption{The area distribution of etch-pit spots of all of $^{56}$Fe ions and ones of secondary fragments on CR-39 surface after CH$_{2}$ target.}
\end{center}
\end{figure}

Figure 2 (a) and (b) show the difference $dx$ and $dy$ in the front surface and back surface on a CR-39 sheet, (c) and (d) show the difference $dx$ and $dy$ before and after the target. If the difference are calculated for all combinations of positions for extracted tracks, only the matching combination ought to make a peak which appears in the figures, and the difference $dx$ and $dy$ of other combinations should be randomly distributed. The deviations $\sigma(x')$ and $\sigma(y')$ give the position accuracies of tracks which are estimated to be $2-4\mu$m for the case of (a) and (b), and $8-20\mu$m for case of (c) and (d).

The matching iron ion track is searched within $4\times\sigma(x')$ and $4\times\sigma(y')$ region of $x'_{th}$ and $y'_{th}$. The matching projectile fragments are searched within the limited fragmentation angle, which is about $10^{\circ}$ in present experiment. The number of projectile fragments leaving the target is determined from the distribution of the etched area. Figure 3 shows the track base area distribution of $^{56}$Fe ions and their fragments in CR-39 sheet. Peaks for $^{56}$Fe and its fragments with charge down to $Z=6$ appear clearly. Because of the limitation of CR-39 detector, the tracks of fragments with charge $1\leq{Z}\leq5$ are not fully etched as a measurable spots. The emission angle $\theta$ of each fragment is calculated by taking readings of the coordinates of the beam track and the fragment track.

\section{Results and discussion}

Emission angular distribution and transverse momentum distribution of projectile fragments provide information on the nuclear structure and the mechanism of nuclear interaction through which fragments are produced. These distributions are also very important in designing experiments with radioactive nuclear beams.

\subsection{Angular distribution}
Emission angle of each fragment and scattering angle of iron ion is calculated from the coordinates of track positions on the surface of CR-39 sheet after the target. The angular uncertainty is determined using the quadruplet fitting method\cite{ref4}
\begin{eqnarray}
\sigma(\theta)=\frac{\sqrt{\sigma^{2}_{z}\sin^{2}\theta+2\sigma^{2}_{p}\cos^{2}\theta}}{2h},
\end{eqnarray}
where $\sigma_{p}$ is the positional uncertainty in $x-y$ plane of the stack coordinate system which is about $3\mu$m for C-target, $\sigma_{z}$ is the positional uncertainty in the z-axis which comes from stack composition and detector thickness measurement and is estimated at $\approx8\mu$m, $\theta$ represents the emission angle of the fitted line. With a detector thickness of $h\approx780\mu$m we thus obtain angular uncertainty $\sigma(\theta)\approx0.16^{\circ}$ for value of $\theta$ up to $8^{\circ}$.

Figure 4 shows the angular distribution of primary iron ions and their fragments for different targets. The emission angle of primary iron ion mainly comes from the Coulomb scattering, most of which is less than $0.6^{\circ}$. Most of projectile fragments have the emission angle less than $1.5^{\circ}$, but some of fragments have emission angle up to $8^{\circ}$. The position of the maximum of the fragment angular distribution increases slightly with increase of the target mass, which can be explained that with the increase of target mass the interaction between projectile and target is increased and influence of fragment suffered from target is increased.
\begin{figure}[htbp]
\begin{center}
\includegraphics[width=0.70\linewidth]{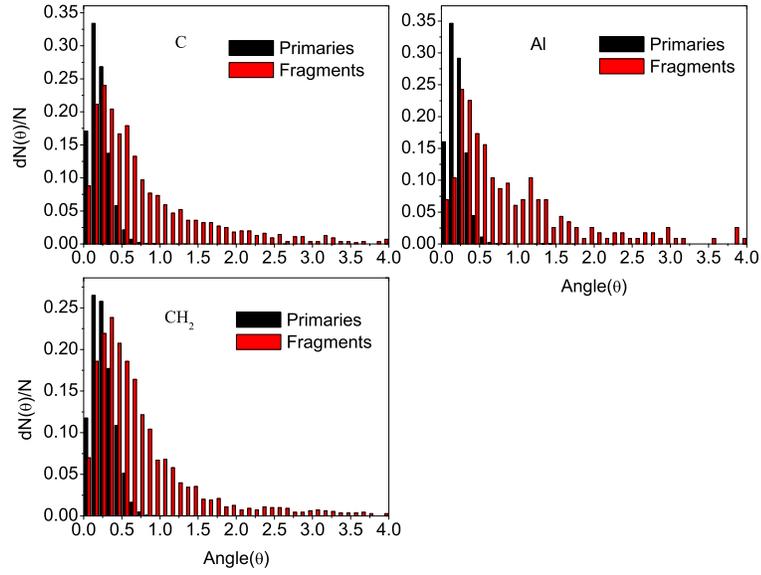}
\caption{The angular distribution of primary iron ions and their fragments for different targets, for comparison the counts of fragments is enlarged two times.}
\end{center}
\end{figure}

\begin{figure}[htbp]
\includegraphics[width=0.49\linewidth]{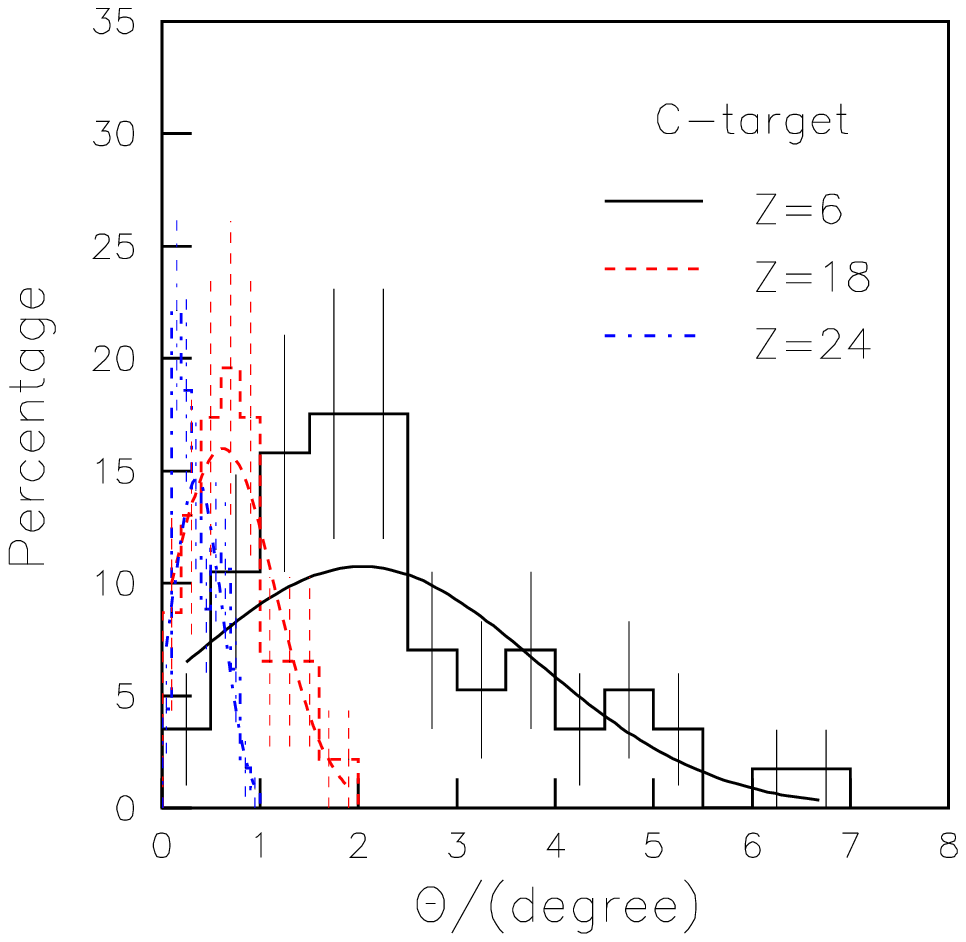}
\includegraphics[width=0.49\linewidth]{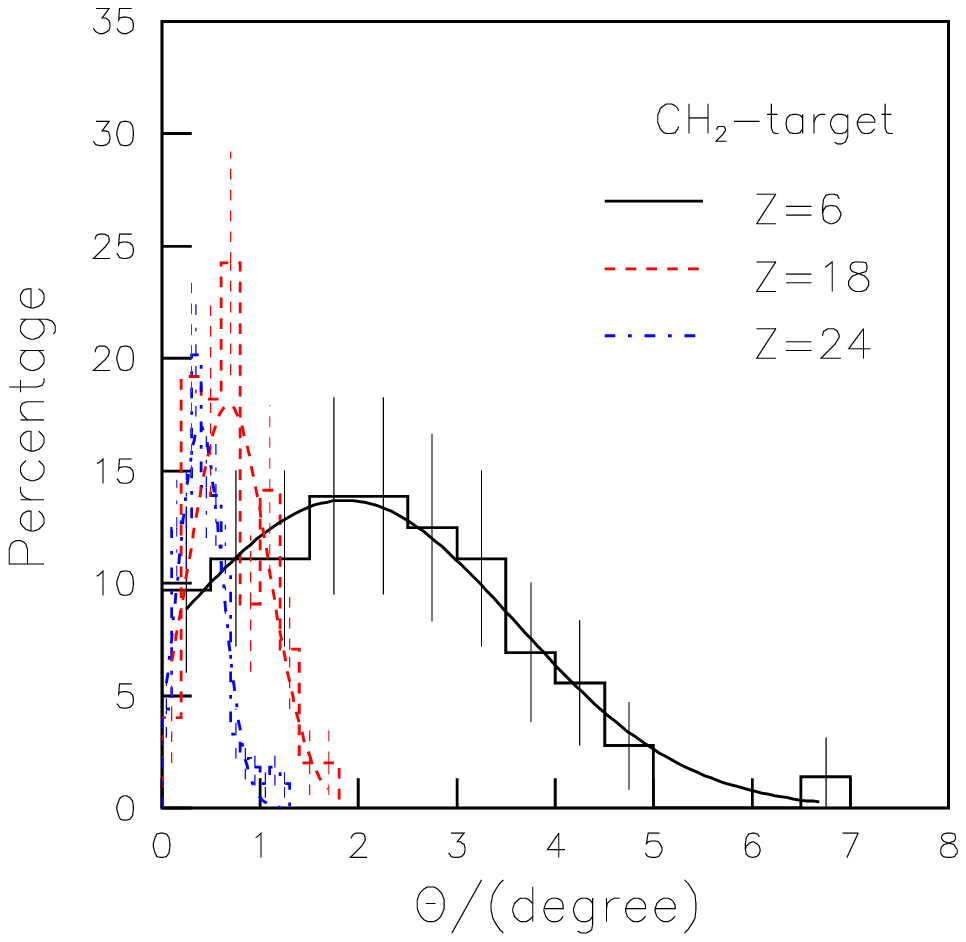}
\caption{ The emission angle distributions of fragments with charge Z=6, 18, and 24, respectively.}
\end{figure}

\begin{figure}[htbp]
\includegraphics[width=0.49\linewidth]{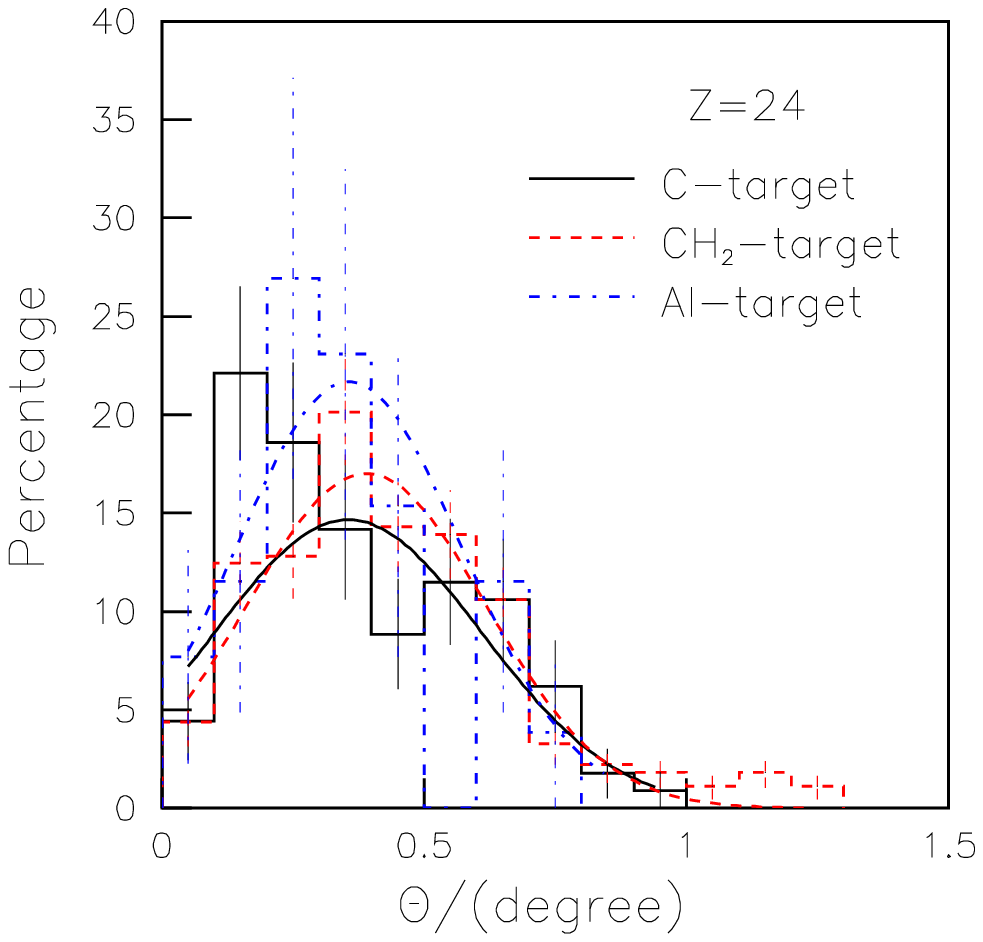}
\includegraphics[width=0.49\linewidth]{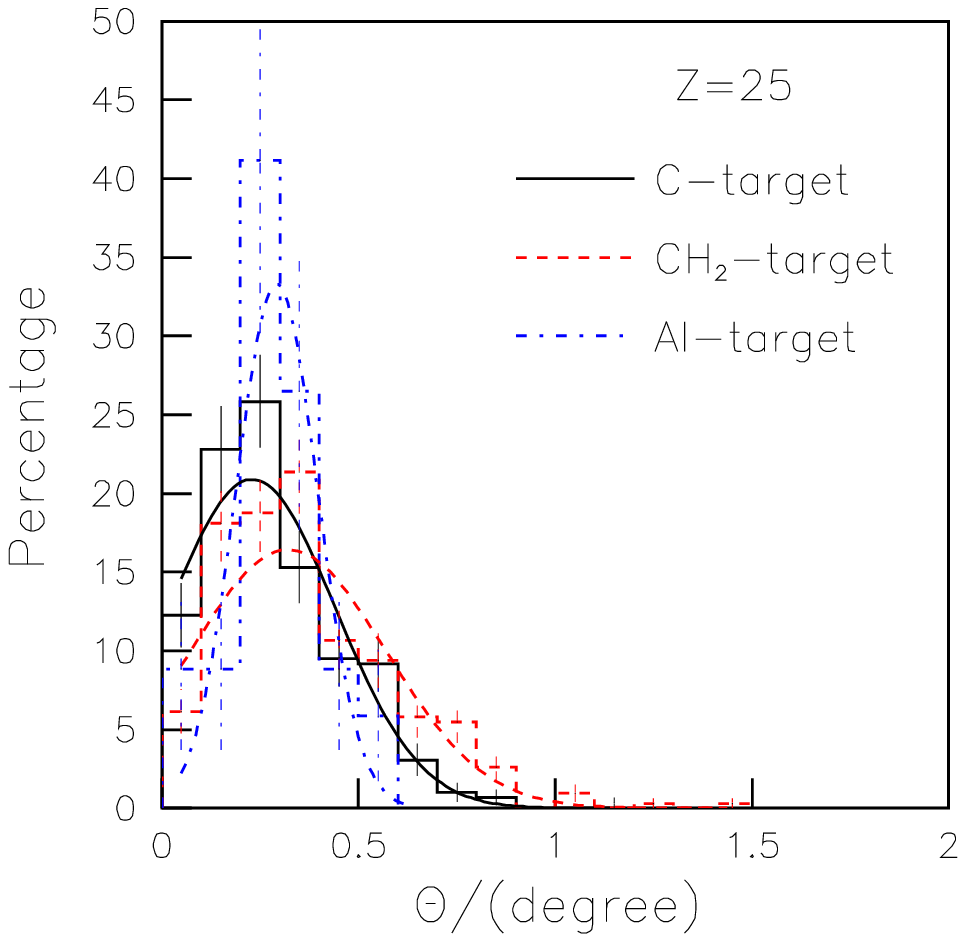}
\caption{ The emission angle distributions of fragments with charge Z=24 and 25, respectively.}
\end{figure}

Fig. 5 shows the emission angle distributions of projectile fragments with $Z=6, 18$, and 24 produced from the fragmentation of $^{56}$Fe on C and CH$_{2}$ targets, and Fig. 6 shows the emission angle distributions of projectile fragments with $Z=24$ and 25 produced from the fragmentation of $^{56}$Fe on C, Al and CH$_{2}$ targets respectively. For fragments from the fragmentation of $^{56}$Fe on C and Al targets, each angular distribution is fitted by a single Gaussian distribution, and for fragments from the fragmentation of $^{56}$Fe on CH$_{2}$ target, each angular distribution is fitted by two Gaussian distribution, the fitting parameters including $\chi^{2}$/DOF are presented in Table 1, where DOF means the degree of freedom of simulation. From Figs. 5 and 6 and results in Table 1 it shows that for the same target the average value and the width of the distribution decrease with increasing the charge number of projectile fragment, and for the same projectile fragment the average value of the distribution increases and the width of the distribution decreases with increasing the target charge number.

\begin{table}
\begin{center}
Table 1. Values of fitting parameters of angular distribution using Gaussian distribution.
\begin{small}
\begin{tabular}{ccccccccc}\hline
Z    & Target    & percentage(1)     & mean value(1)    & error(1)         & percentage(2)    & mean value(2) &  error(2)  & $\chi^{2}$/DOF  \\\hline
 25   & CH$_{2}$  & $16.38\pm5.52$  & $0.25\pm0.02$     & $0.13\pm0.03$       & $7.33\pm2.92$  & $0.49\pm0.17$  & $0.28\pm0.08$  & 1.12   \\
      & C         & $20.89\pm1.62$  & $0.23\pm0.02$     & $0.21\pm0.02$       &                  &                  &            & 1.81   \\
      & Al        & $33.28\pm13.27$ & $0.29\pm0.02$     & $0.10\pm0.04$       &                  &                  &            & 1.52   \\
 24   & CH$_{2}$  & $17.73\pm1.50$  & $0.39\pm0.02$     & $0.21\pm0.02$       & $1.44\pm0.54$  & $1.13\pm0.11$  & $0.18\pm0.16$  & 1.15  \\
      & C         & $14.66\pm1.84$  & $0.36\pm0.03$     & $0.26\pm0.03$       &                  &                  &            & 2.15   \\
      & Al        & $21.68\pm5.69$  & $0.36\pm0.05$     & $0.22\pm0.05$       &                  &                  &            & 0.31  \\
 18   & CH$_{2}$  & $22.90\pm3.61$  & $0.53\pm0.06$     & $0.25\pm0.04$       & $10.93\pm4.26$ & $1.16\pm0.08$  & $0.18\pm0.08$  & 2.17   \\
      & C         & $16.00\pm3.58$  & $0.62\pm0.12$     & $0.54\pm0.14$       &                  &                  &            & 0.40  \\
 6    & CH$_{2}$  & $13.71\pm0.95$  & $1.86\pm0.34$     & $1.73\pm0.31$       & $0.30\pm1.06$  & $2.97\pm1.07$  & $0.04\pm1.44$  & 0.12  \\
      & C         & $10.74\pm3.66$  & $2.04\pm0.34$     & $1.78\pm0.81$       &                  &                  &            & 1.01 \\\hline
\end{tabular}
\end{small}
\end{center}
\end{table}

\subsection{Transverse momentum distribution}
The transverse momentum per nucleon ($p_{t}$) of a projectile fragment was calculated on the basis of its emission angle $\theta$,
\begin{equation}
p_{t}=p\sin\theta,
\end{equation}
where p is the momentum per nucleon of beam which can be calculated from beam energy per nucleon (E), $p=(E^{2}+2m_{0}E)^{1/2}$. $m_{0}$ is the nucleon rest mass and $\theta$ the emission angle of the projectile fragment with respect to the beam direction.

Figure 7 shows the transverse momentum distribution of projectile fragment with charge $Z=6, 18$, and 24 produced from the fragmentation of $^{56}$Fe on C and CH$_{2}$ targets. The distributions can be well fitted by a single Gaussian distribution, which is the same as the distributions of fragments produced in reactions of light projectiles\cite{ref28, ref29} and heavy projectiles\cite{ref30,ref31,ref32,ref33,ref34} at relativistic energies. These Gaussians are in good agreement with predictions of the statistical model of Goldhaber\cite{ref35}. This model assumes that the Fermi momenta of the nucleons in a fragment are statistically distributed as those in the original projectile nucleus. The averaged transverse momentum per nucleon for fragment with charge $Z=6, 18$, and 24 are $35.89\pm35.25$, $11.34\pm8.74$, and $6.91\pm4.74$ A MeV/c, respectively, for C target, and $34.38\pm30.79$, $11.98\pm7.73$, and $7.08\pm4.25$ A MeV/c, respectively, for CH$_{2}$ target. For the same target the averaged transverse momentum per nucleon, and the width of the distribution increase with the decrease of the charge of fragment. This tendency is also observed in Ref.\cite{ref34}. 

Projectile fragments come from the directly produced fragments (primary fragments) and the sequential decay fragments from excited primary fragment. However, since the primary fragments are excited, they are deexcited by light particle evaporation. This secondary decay decreases the observed masses and increase the observed momentum widths of the primary fragments\cite{ref36}. The contribution from sequential decay of primary fragments to the heavy projectile fragments are less than that to the light projectile fragments, the widths of the transverse momentum distributions of light projectile fragments are greater than that of the heavy projectile fragments. So the transverse momentum distribution width deceases with the increase of the charge of the projectile fragment, which is shown in Figure 7.

Figure 8 shows relation of the averaged transverse momentum per nucleon and the charge of projectile fragment for interactions of $^{56}$Fe and C, Al, and CH$_{2}$ targets. The averaged transverse momentum per nucleon decreases with the increase of fragment size, no obviously target size dependence is observed in present investigation. Figure 9 shows the transverse momentum distribution of all fragments for interactions of $^{56}$Fe and C, Al, and CH$_{2}$ targets, no obviously target size dependence is also observed. The heavy fragment comes from peripheral collisions, that is collisions with larger impact parameter. The light fragment comes from central and semi-central collisions, that is collisions with smaller impact parameter. According to the participant-spectator model\cite{ref37}, with the increase of impact parameter the overlapped region decreases, the communication between participant and spectator decrease. This results in the decrease of the excitation energy of projectile fragments, so the average of transverse momentum of fragment is decreased.
\begin{figure}[th]
\includegraphics[width=0.49\linewidth]{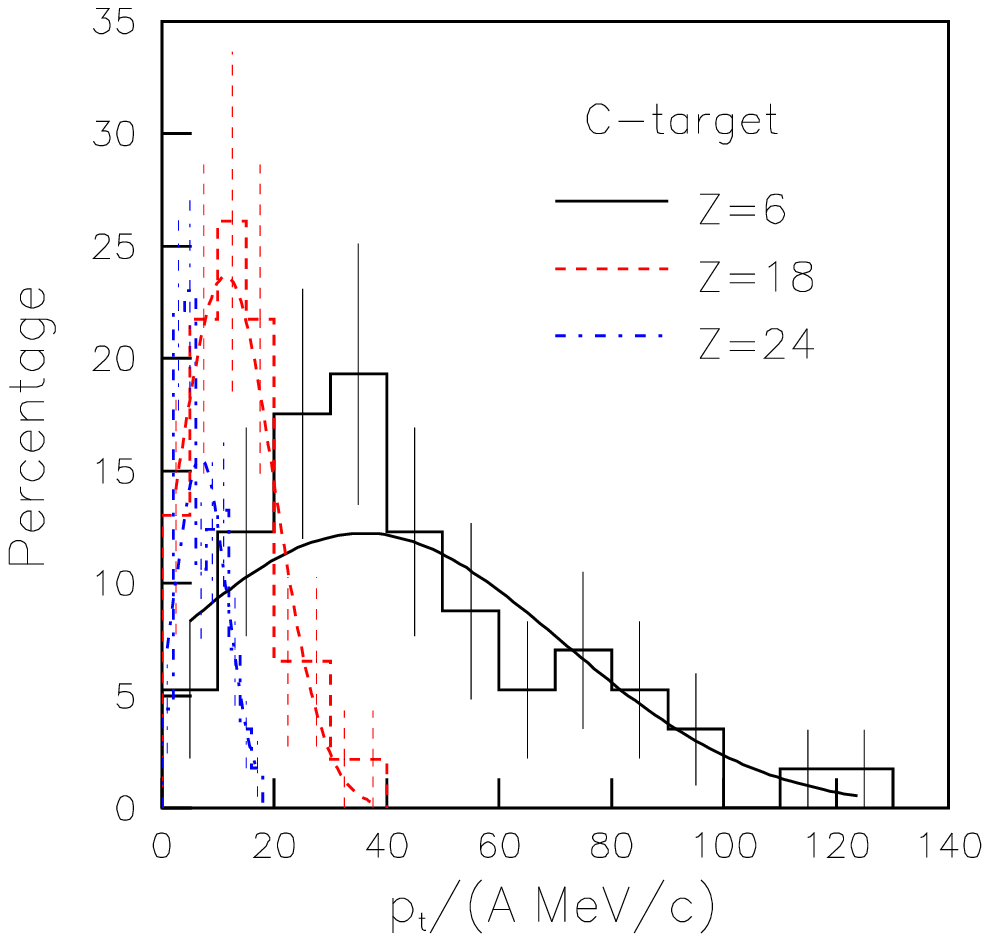}
\includegraphics[width=0.49\linewidth]{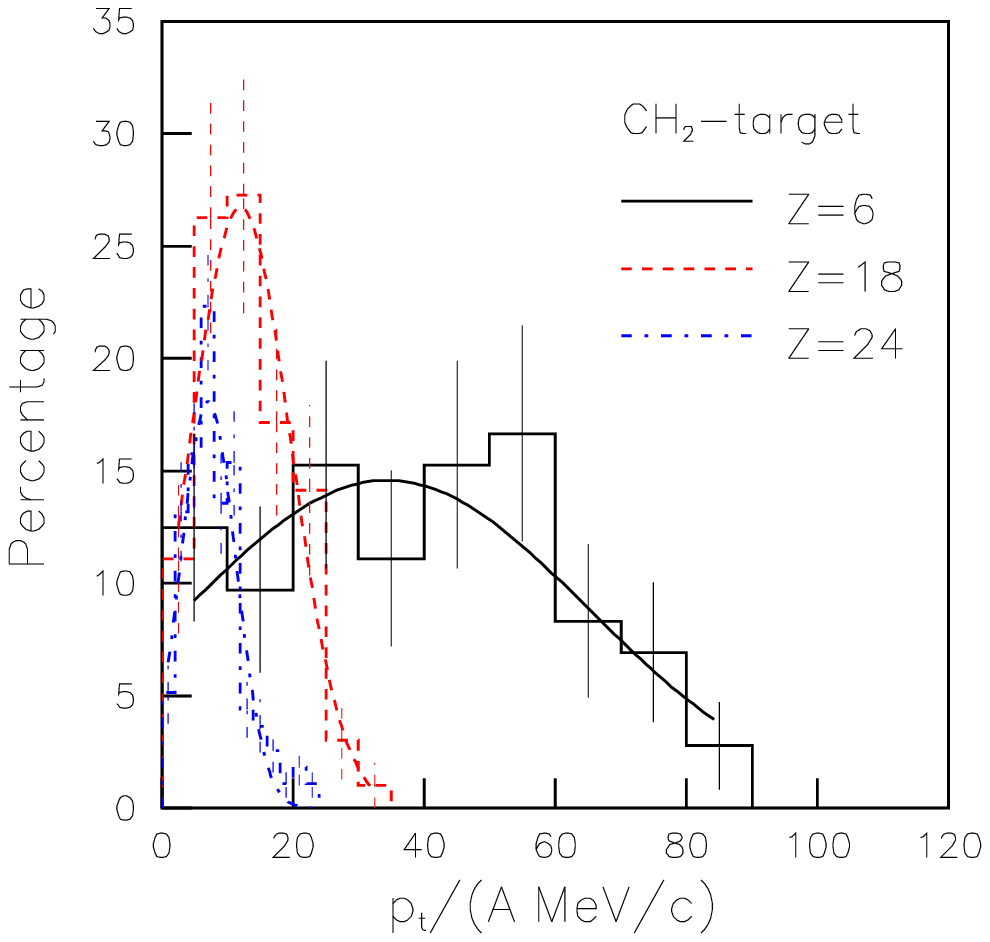}
\caption{The transverse momentum distributions of fragments with charge Z=6, 18, and 24, respectively.}
\end{figure}

\begin{figure}[htbp]
\begin{center}
\includegraphics[width=0.65\linewidth]{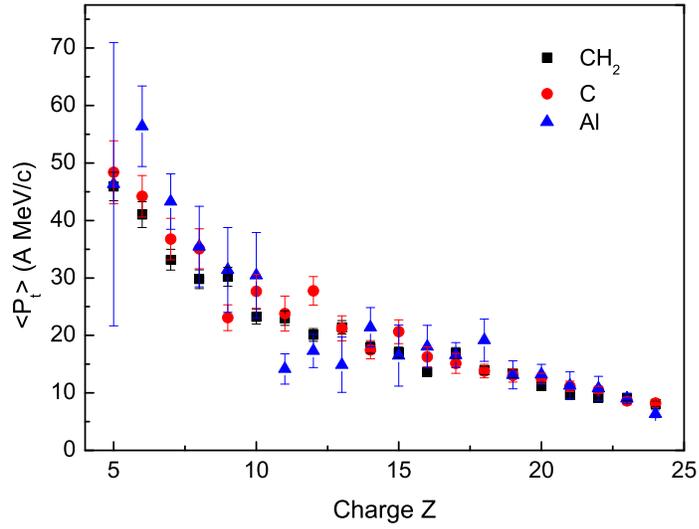}
\caption{Relation of the averaged transverse momentum and the charge of projectile fragment for interactions of $^{56}$Fe and C, Al, and CH$_{2}$ targets.}
\end{center}
\end{figure}

\begin{figure}[htbp]
\begin{center}
\includegraphics[width=0.65\linewidth]{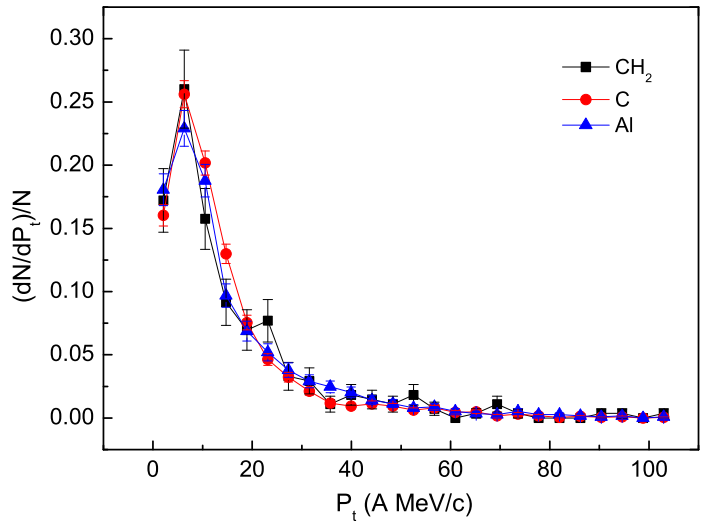}
\caption{Transverse momentum distribution of all fragments for interactions of $^{56}$Fe and C, Al, and CH$_{2}$ targets.}
\end{center}
\end{figure}

\begin{figure}[htbp]
\includegraphics[width=0.49\linewidth]{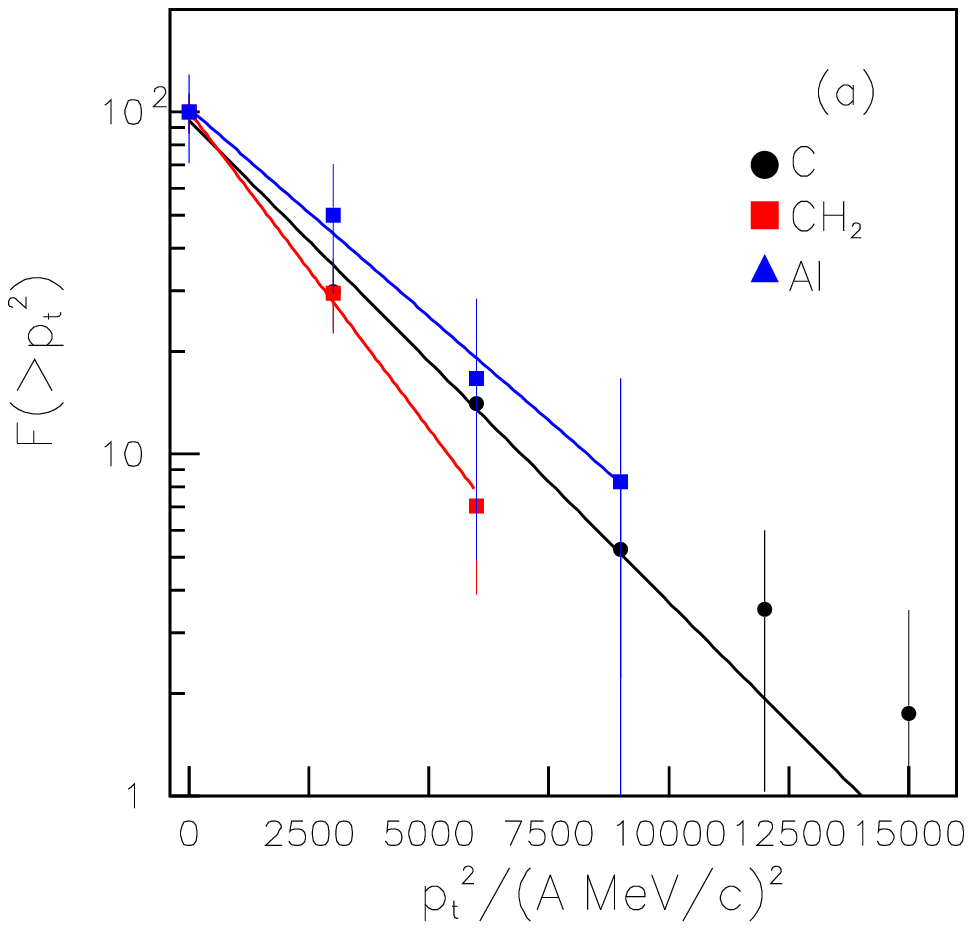}
\includegraphics[width=0.49\linewidth]{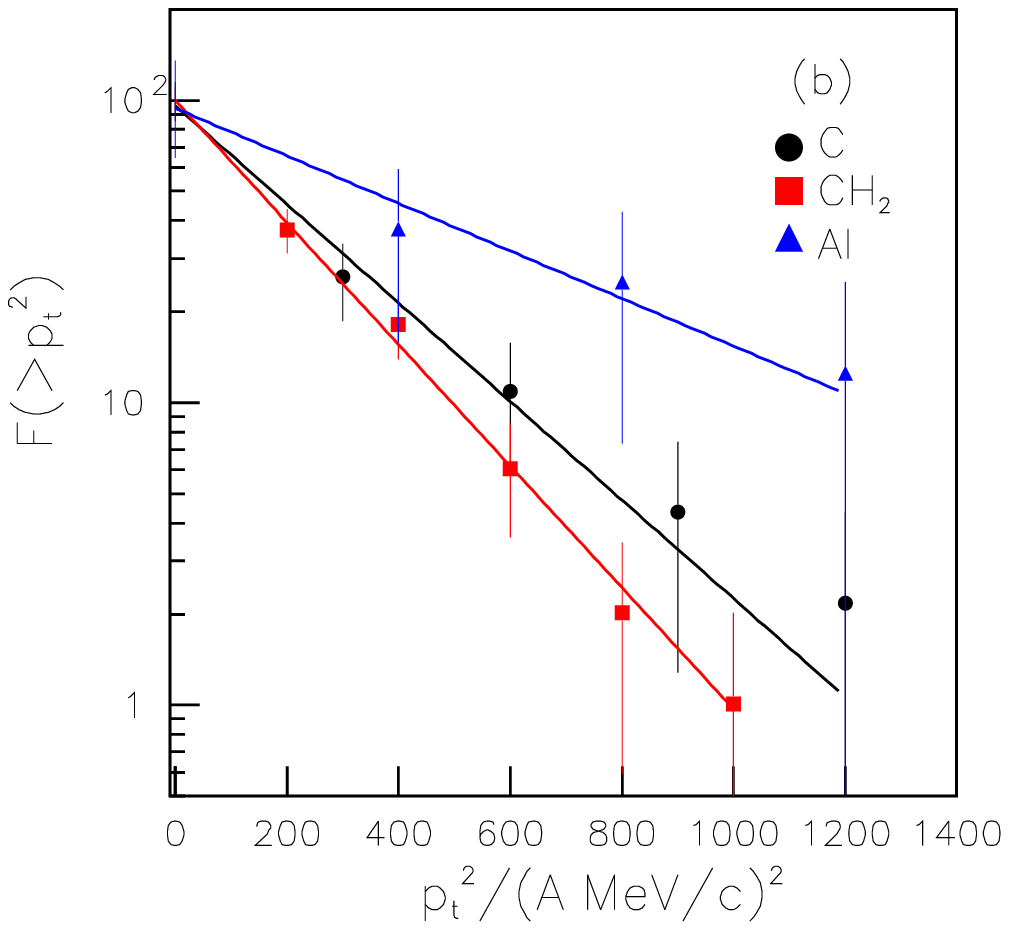}
\includegraphics[width=0.49\linewidth]{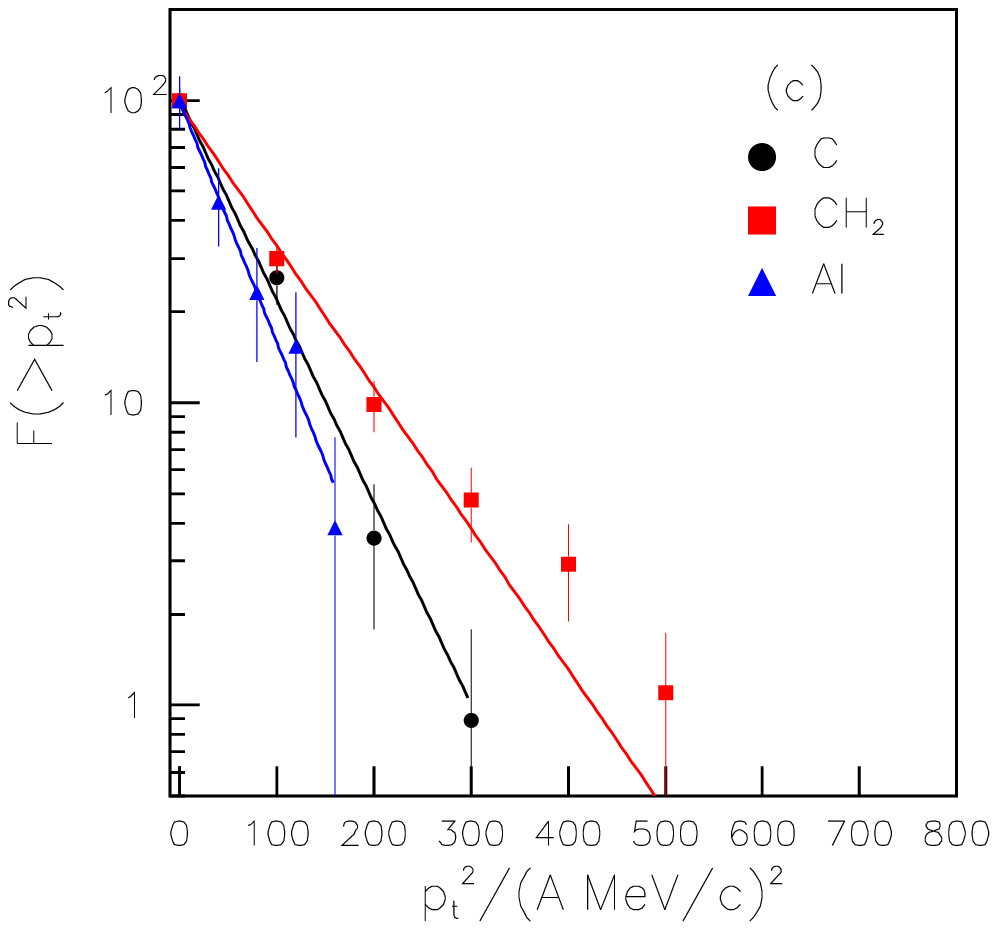}
\includegraphics[width=0.49\linewidth]{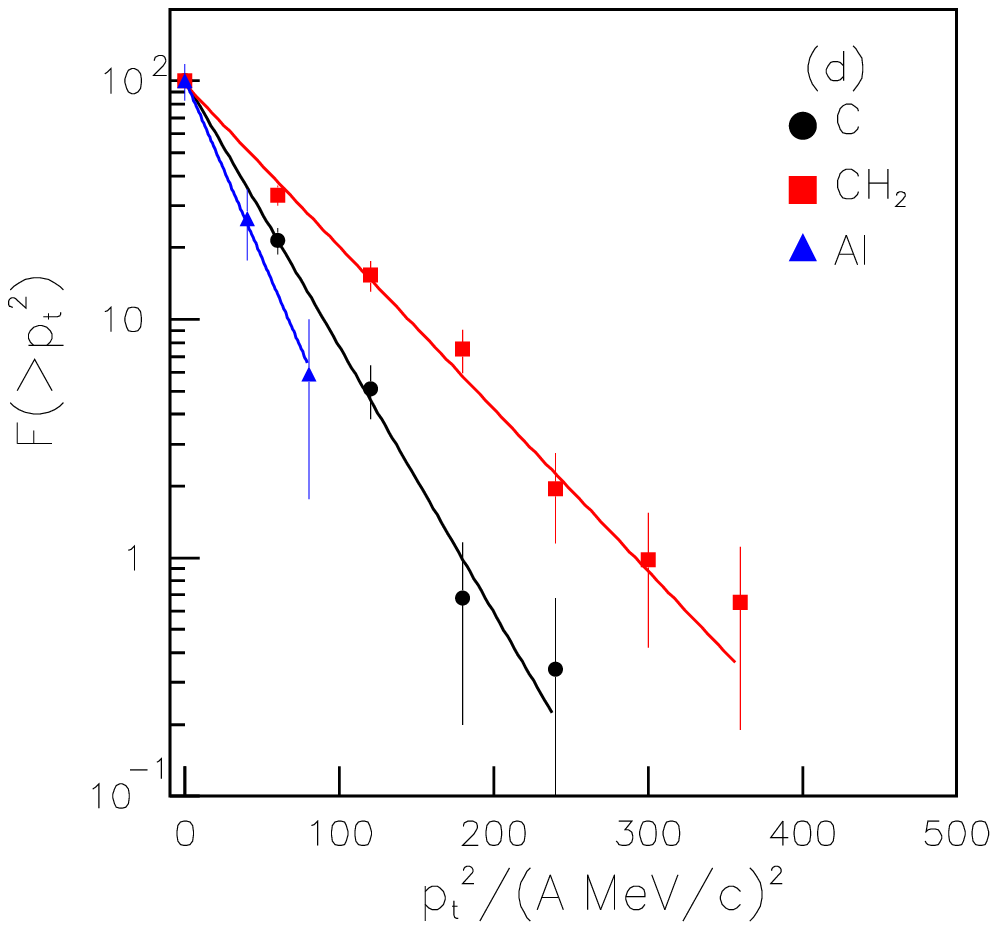}
\caption{The cumulative $p_{t}^{2}$ distribution of projectile fragments with charge $Z=6$ (a), $Z=18$ (b), $Z=24$ (c), and $Z=25$ (d).}
\end{figure}

Based on the participant-spectator concept and the fireball model\cite{ref38}, the large number of swept out nucleons combined with an anticipated fairly large number of interactions per particle is presumably responsible for the quasi-equilibrated system, {\it i.e.} the fireball which can then be described in term of mean value and statistical (Maxwell-Boltzmann) distribution. If we assume that the emission of projectile fragments is Maxwell-Boltmann distribution in the projectile rest frame with a certain temperature T, then the integral frequency distribution of the square of the transverse momentum per nucleon is
\begin{equation}
lnF(>p_{t}^{2})=-\frac{A}{2M_{p}T}p_{t}^{2}
\end{equation}
where A is the mass number of fragment, $M_{p}$ is the mass of proton. The linearity of such a plot would be strong evidence for a single temperature of emission source.

Figure 10 shows the cumulative plots of $F(>p^{2}_{t})$ as a function of $p^{2}_{t}$ for projectile fragments from the fragmentation of $^{56}$Fe on C, Al and CH$_{2}$ targets. All of the plots can be well fitted by a single Rayleigh distribution of the form
\begin{equation}
F(p_{t}^{2})=Cexp(\frac{-p_{t}^{2}}{2\sigma^{2}}),
\end{equation}
where $\sigma=\sqrt{2/\pi}<p_{t}>$, which is related to the temperature of fragment emission source, $T=\sigma^{2}A/M_{p}$. The fitting parameters including $\chi^{2}$/DOF and the temperature of the emission source are presented in Table 2. Because CR-39 detector can not identify the mass numbers of projectile fragments, we use the mass number of stable nucleus to calculate the temperature of projectile fragment emission source. The influence from isotope is less than $1\%$ when the abundance of isotope is considered. The dependence of the temperature of emission source on the size of fragment for different targets is shown in figure 11. From the results of Table 2 and figure 11 we can conclude that the temperature of projectile fragment emission source decreases with the increase of the charge of fragment for the same targets. The temperature increases with the increase of target size for emission of fragment with change $Z=6$ and 18, but for emission of fragment with change $Z=24$ and 25 this dependence is not obvious.

\begin{table}
\begin{center}
Table 2. Values of fitting parameters of $p_{t}^{2}$ distribution using Rayleigh distribution and the temperature of the emission source.
\begin{tabular}{cccccc}\hline
 Z     & Target   & C                   & $\sigma^{2}$ ((MeV/c)$^{2}$)  &   T (MeV)     &   $\chi^{2}$/DOF\\\hline
 25   & CH$_{2}$  & $96.52\pm5.35$      & $31.93\pm1.69$         & $1.87\pm0.10$ &  0.831   \\
      & C         & $100.16\pm5.71$     & $19.47\pm1.11$         & $1.14\pm0.07$ &  0.235 \\
      & Al        & $100.56\pm16.91$    & $14.50\pm2.78$         & $0.85\pm0.16$ &  0.033  \\
 24   & CH$_{2}$  & $96.77\pm5.95$      & $46.51\pm3.01$         & $2.58\pm0.17$ &  1.398   \\
      & C         & $101.72\pm9.24$     & $32.52\pm2.89$         & $1.80\pm0.16$ &  0.575 \\
      & Al        & $99.56\pm17.90$     & $27.25\pm5.00$         & $1.51\pm0.28$ &  0.159  \\
 18   & CH$_{2}$  & $99.45\pm9.32$      & $108.00\pm9.34$        & $4.61\pm0.40$ &  0.142   \\
      & C         & $95.86\pm14.60$     & $133.30\pm22.31$       & $5.68\pm0.95$ &  0.310 \\
      & Al        & $94.08\pm33.96$     & $276.24\pm127.42$      & $11.78\pm5.43$ & 0.108  \\
 6    & CH$_{2}$  & $100.76\pm11.64$    & $1168.60\pm156.12$     & $14.95\pm2.00$ & 0.119  \\
      & C         & $94.57\pm12.88$     & $1541.40\pm221.88$     & $19.72\pm2.84$ & 0.400 \\
      & Al        & $102.55\pm27.41$    & $1783.50\pm523.46$     & $22.82\pm6.70$ & 0.065  \\\hline
\end{tabular}
\end{center}
\end{table}

\begin{figure}[htbp]
\begin{center}
\includegraphics[width=0.65\linewidth]{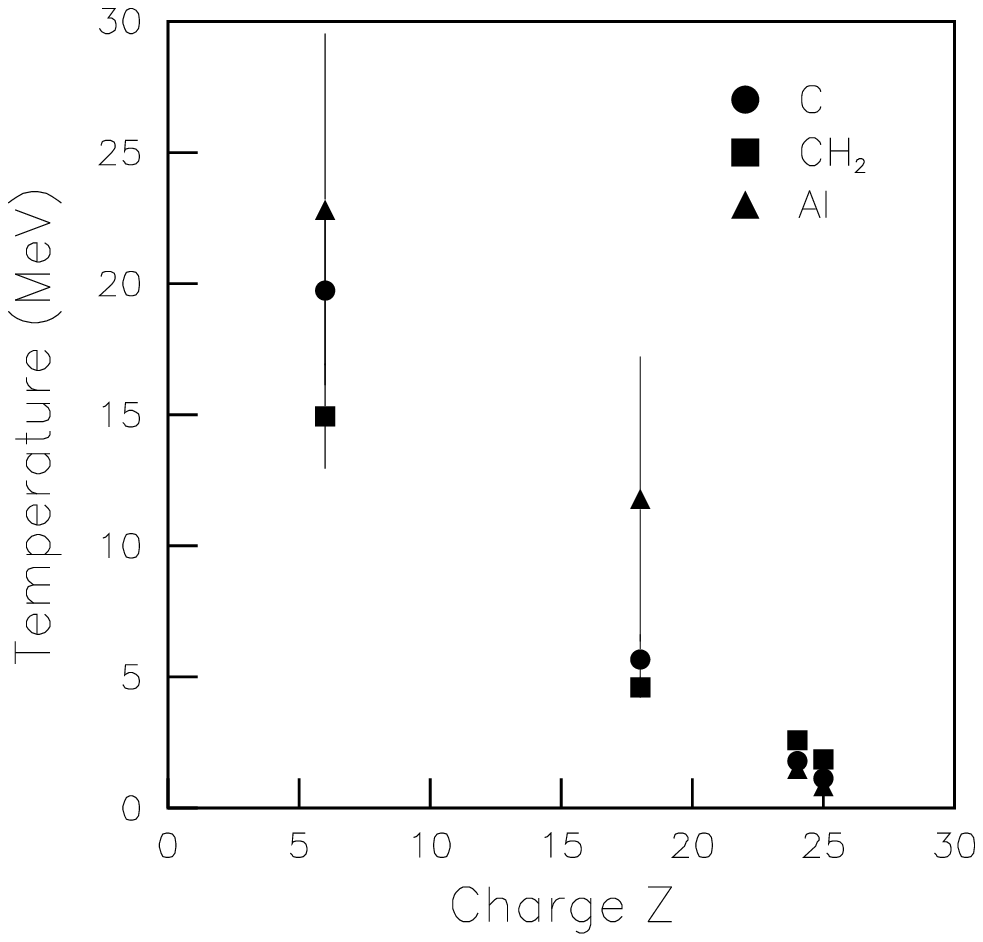}
\caption{The temperature derived from the distribution of $p_{T}^{2}$.}
\end{center}
\end{figure}

The temperature of projectile fragment emission source have been investigated by different collaborations[30,31,39-44]. ALADIN Collaboration studied the slope temperature ($T_{slope}$) for spectator decay as a function of the fragment mass ($A_{frag}$) for spectator decays following $^{197}$Au on $^{197}$Au collisions, they found that there is a rapid increase of $T_{slope}$ with fragment mass which saturates for $A_{frag}\geq3$ around $T_{slope}\sim17$ MeV\cite{ref39,ref45}. EOS Collaboration also studied the variation of remnant temperature with the charged particle multiplicity, they found that the remnant temperature increases with increase of the charged particle multiplicity and the maximum is about $15.6\pm0.47$ MeV\cite{ref46}. These maximum temperature is the same as our results within  experimental errors for emission of fragment with charge $Z=6$.

According to the participant-spectator concept, it is assumed that when the interaction of projectile and target nuclei takes place, the projectile and target sweep out cylindrical cuts through each other. During the separation of the spectators from the participants, there is some intercommunication, which results in the excitation of the spectators. This excitation strongly depends on the contacted area of the colliding system. The heavier fragment is corresponding to the large impact parameter and small contacted areas, the lighter fragment is corresponding to the smaller impact parameter and large contacted areas. So the excitation energy of the heavier fragment is less than that of the lighter fragment, which results in the temperature of emission source of heavier fragment is less than that of the lighter fragment.

\section{Summary}
The emission angular distribution and the transverse momentum distribution of projectile fragments produced in fragmentation of $^{56}$Fe on C, Al, and CH$_{2}$ targets are studied in present investigation. It is found that for the same target the average value and width of angular distribution decrease with increase of the projectile fragment charge, and for the same projectile fragment the average value of the distribution increases and the width of the distribution decreases with increasing the target charge number. The transverse momentum distribution of projectile fragment can be explained by a single Gaussian distribution and the averaged transverse momentum per nucleon decreases with the increase of the charge of fragment, and no obvious dependence of transverse momentum on target size is observed. The cumulated squared transverse momentum distribution of fragment can be well explained by a single Rayleigh distribution. The temperature parameter of emission source of projectile fragments decreases with the increase of the size of projectile fragments.

\section{Acknowledgements}
This work has been supported by the Chinese National Science Foundation under Grant Nos: 11075100 and 10975019 and the Natural Foundation of Shanxi Province under Grant 2011011001-2, the Shanxi Provincial Foundation for Returned Overseas Chinese Scholars, China (Grant No. 2011-058), the Foundation of Ministry of Personnel of China for Returned Scholars Grant MOP2006138, and the Fundamental Research Funds for the Central Universities. We gratefully acknowledge the staffs of the HIMAC for providing the beam to expose the stacks.

\end{document}